\documentclass[aps,prl,twocolumn,superscriptaddress,groupedaddress]{revtex4}
\usepackage{amssymb}
\usepackage{cancel}
\usepackage{color,graphicx}
\usepackage{amsmath}
\usepackage{amsbsy}
\usepackage{amsthm}
\usepackage{bbm}
\usepackage{epsfig}
\usepackage{lscape}
\usepackage{float}
\usepackage{graphicx}
\usepackage{subfigure}
\usepackage{dcolumn}
\usepackage{bbm}
\usepackage{color,epstopdf}
\usepackage{amscd}
\usepackage{amsfonts}  
\usepackage[]{amsmath}
\usepackage{amssymb}    
\usepackage{mathrsfs}
\usepackage{verbatim}
\usepackage[]{cases}
\usepackage{amsmath}
\usepackage{wasysym}
\usepackage[utf8]{inputenc}
\usepackage[T1]{fontenc}
\usepackage{mathtools}
\usepackage[dvipsnames]{xcolor}
\usepackage{url}

\usepackage[colorlinks=true,linkcolor=blue,urlcolor=blue,citecolor=blue,pdfusetitle]{hyperref}
\usepackage{soul}

\newcommand{\ket}[1]{\left| {#1} \right\rangle}
\newcommand{\bra}[1]{\left\langle {#1}\right|}

\newcommand{\be}{\begin{equation}}
\newcommand{\ee}{\end{equation}}

\newcommand{\iu}{\mathrm{i}} 
\newcommand{\ec}{\mathrm{e}} 

\begin{document}

\title{Quantum Temporal Superposition: the case of QFT}

\author{Laura J. Henderson}
\email[]{l7henderson@uwaterloo.ca}
\affiliation{Department of Physics and Astronomy, University of Waterloo, Waterloo, Ontario, Canada, N2L 3G1}
\affiliation{Institute for Quantum Computing, University of Waterloo, Waterloo, Ontario, Canada, N2L 3G1}
\affiliation{Centre for Quantum Computation and Communication Technology,School of Science, RMIT University, Melbourne, Victoria 3001, Australia}
\author{Alessio Belenchia}
\email[]{a.belenchia@qub.ac.uk}
\affiliation{Centre for Theoretical Atomic, Molecular, and Optical Physics,
School of Mathematics and Physics, Queens University, Belfast BT7 1NN, United Kingdom}
\author{Esteban Castro-Ruiz}
\email[]{esteban.castro.ruiz@univie.ac.at}
\affiliation{QuIC, Ecole polytechnique de Bruxelles, C.P. 165, Universit\'e libre de Bruxelles, 1050 Brussels, Belgium}
\affiliation{Faculty of Physics, University of Vienna, Boltzmanngasse 5, 1090 Vienna}
\affiliation{Institute for Quantum Optics and Quantum Information (IQOQI), Boltzmanngasse 3 1090 Vienna, Austria}
\author{Costantino Budroni}
\email[]{costantino.budroni@univie.ac.at}
\affiliation{Faculty of Physics, University of Vienna, Boltzmanngasse 5, 1090 Vienna}
\affiliation{Institute for Quantum Optics and Quantum Information (IQOQI), Boltzmanngasse 3 1090 Vienna, Austria}
\author{Magdalena Zych}
\email[]{m.zych@uq.edu.au}
\affiliation{Centre for Engineered Quantum Systems,
School of Mathematics and Physics,
The University of Queensland,
St Lucia, QLD 4072, Australia}
\author{\v{C}aslav Brukner}
\email[]{caslav.brukner@univie.ac.at}
\affiliation{Faculty of Physics, University of Vienna, Boltzmanngasse 5, 1090 Vienna}
\affiliation{Institute for Quantum Optics and Quantum Information (IQOQI), Boltzmanngasse 3 1090 Vienna, Austria}
\author{Robert B. Mann}
\email[]{rbmann@uwaterloo.ca}
\affiliation{Department of Physics and Astronomy, University of Waterloo, Waterloo, Ontario,  Canada, N2L 3G1}
\affiliation{Institute for Quantum Computing, University of Waterloo, Waterloo, Ontario, Canada, N2L 3G1}

\begin{abstract}
Quantum field theory is completely characterized by the field correlations between spacetime points. In turn, some of these can be accessed by locally coupling to the field simple quantum systems, a.k.a. particle detectors. In this work, we consider what happens when a quantum-controlled superposition of detectors at different space-time points is used to probe the correlations of the field. We show that, due to quantum interference effects, two detectors can gain information on field correlations which would not be otherwise accessible. This has relevant consequences for information theoretic quantities, like entanglement and mutual information harvested from the field. In particular, the quantum control allows for extraction of  entanglement in scenarios where this is otherwise provably impossible.
\end{abstract}
\maketitle

\noindent\emph{Introduction---} The relationship between effects and causes is foundational to our everyday experience and is rooted in the basic laws of physics as we presently understand them.  Both in classical and quantum physics, events happen in a  fixed causal order.   

One of the more intriguing proposals put forward in recent years, is that quantum physics may admit nonclassical causal structures where the order of events is indefinite
\cite{2007JPhA...40.3081H,chiribella2012perfect,2013PhRvA..88b2318C,2012NatCo...3.1092O}.  Apart from advantages afforded in computation 
\cite{2014PhRvL.113y0402A,Chiribella-2019NatCom}
and communication~\cite{2015PhRvA..92e2326F,PhysRevLett.117.100502,PhysRevLett.120.120502,2018arXiv180906655S,PhysRevA.99.062317}, the  incorporation of this idea into the foundations of physics may provide new insights toward realizing a quantum theory of gravity
\cite{2007JPhA...40.3081H,Zych:2017tau,paunkovic2019causal}.   This idea has become known as ``Indefinite Causal Order", or ICO.

Recently a scenario featuring an indefinite causal order of events has been experimentally realized in the form of a quantum switch~\cite{2015NatCo...6.7913P,Rubinoe1602589,2018PhRvL.121i0503G}.  In this setting the order in which two quantum operations $A$ and $B$
are performed on some target is coherently controlled by a quantum system, placing the order of
operations in temporal superposition~\cite{PhysRevLett.64.2965}. 
  An ordinary quantum circuit using the same number of operations cannot reproduce the effects of the quantum switch~\cite{chiribella2012perfect,2014PhRvL.113y0402A}.

 Little is known about how ICO affects quantum entanglement, particularly entanglement of quantum fields. The vacuum state of a free quantum field has long been known to  contain correlations between timelike and spacelike separated regions~\cite{summers1985bell,summers_bells_1987}, and it is of foundational interest to understand how such correlations are affected by ICO.  Vacuum correlations are at the core of the black hole information paradox~\cite{Preskill:1992tc,Mathur:2009hf} and its proposed solutions~\cite{Almheiri:2012rt,Braunstein:2009my,Mann:2015luq}, and play a key role in quantum energy teleportation~\cite{doi:10.1143/JPSJ.78.034001,Hotta:2011xj}. 
 
 Furthermore, this vacuum entanglement can be extracted from the field by swapping it with (originally separable) detectors.  This procedure is called Entanglement Harvesting~\cite{valentini1991non,reznik2003entanglement,PhysRevA.71.042104,Salton:2014jaa}, and is most easily  analysed by considering the detectors to be 2-level systems (qubits) each with a ground and excited state. These objects are called Unruh-de Witt (UDW) detectors.

In relativistic physics, the causal order between events is constrained by the light cones at each space-time point. In the case in which we probe the field via UDW detectors, their interaction with the field constitutes an event. Traditionally, UDW detectors are assumed to probe the quantum field at fixed space-time points, so that all events happen in a definite causal order. Our ability to, e.g., harvest entanglement from the vacuum is constrained by this fact. 

In this letter, we consider what happens when a quantum-controlled  superposition of two UDW detectors is used to probe the correlations of a scalar field vacuum. Specifically, a single qubit controls the temporal switching of both detectors, which are assumed to follow inertial trajectories. We identify two main cases of interest for a  quantum control: space-like regions in superposition of different times, and time-like regions in superposition of different causal orders (ICO).
We find in both cases that, thanks to quantum interference effects, the two detectors can gain information about field correlations that is impossible to access with only a single use of each detector in the standard scheme with no coherent control. This is shown to have relevant consequences for information theoretic quantities, like entanglement and mutual information harvested by the detectors from the field.    
\\
\\
\noindent\emph{Quantum Control of two UDW Detectors.--}
The Hamiltonian of an UDW detector coupled to a real scalar field --- a simplified version of the electric dipole light-matter interaction --- is given in full generality by
\begin{equation}
    H_I(t)=\lambda_D\int d\mathbf{x}S(\mathbf{x})\chi_{D}(t)(e^{i\Omega_D t}\sigma_+ +e^{-i\Omega_D t}\sigma_{-})\otimes\phi(x_{D}(t)),
\end{equation}
where $\Omega_D$ is the energy-gap of the detector, $\lambda_{D}$ its coupling constant, $S(\mathbf{x})$ is a smearing function  that models the spatial extent of the detector,
 and $\chi_D(t)$ is a window function that determines the spacetime activation regions, i.e.
 when the detector is on or off~\footnote{See~\cite{stritzelberger2019coherent} for the recent inclusion of the effect of the coherent spreading of the center-of-mass wave-function of an UDW detector on its interaction with the field.}.  
 The $\sigma_{\pm}$ are the ladder operators of the two-level UDW detector and $\phi$ is the scalar field. Here and in the following we consider stationary detectors, so that $x_D(t)=x_D=\rm{const.}$

We  consider a composite system formed by two {pointlike} UDW detectors, locally coupled to a real scalar field, whose Hamiltonian is 
\begin{align}
   H_I(t) & =\sum_{D=A,B}\sum_{i=0,1}\lambda_D\,\chi_{D,i}(t)(e^{i\Omega_D t}\sigma_+ +e^{-i\Omega_D t}\sigma_{-}) \nonumber\\
   & \qquad \otimes\phi(x_{D}(t))\otimes\ket{i}_C\bra{i}
\label{Ham}   
\end{align}
where  $i=0,1$ is determined by the computational basis states $\{\ket{0},\ket{1}\}$ of a control qubit 
(system $C$ henceforward) that
governs the window functions of the  detectors  $A$ and $B$. Placing the initial state of the qubit in a coherent superposition of the computational basis states will thus  give rise to a superposition of different switching times of the two detectors. 

\begin{figure}[h!]
\centering
\includegraphics[width=\columnwidth]{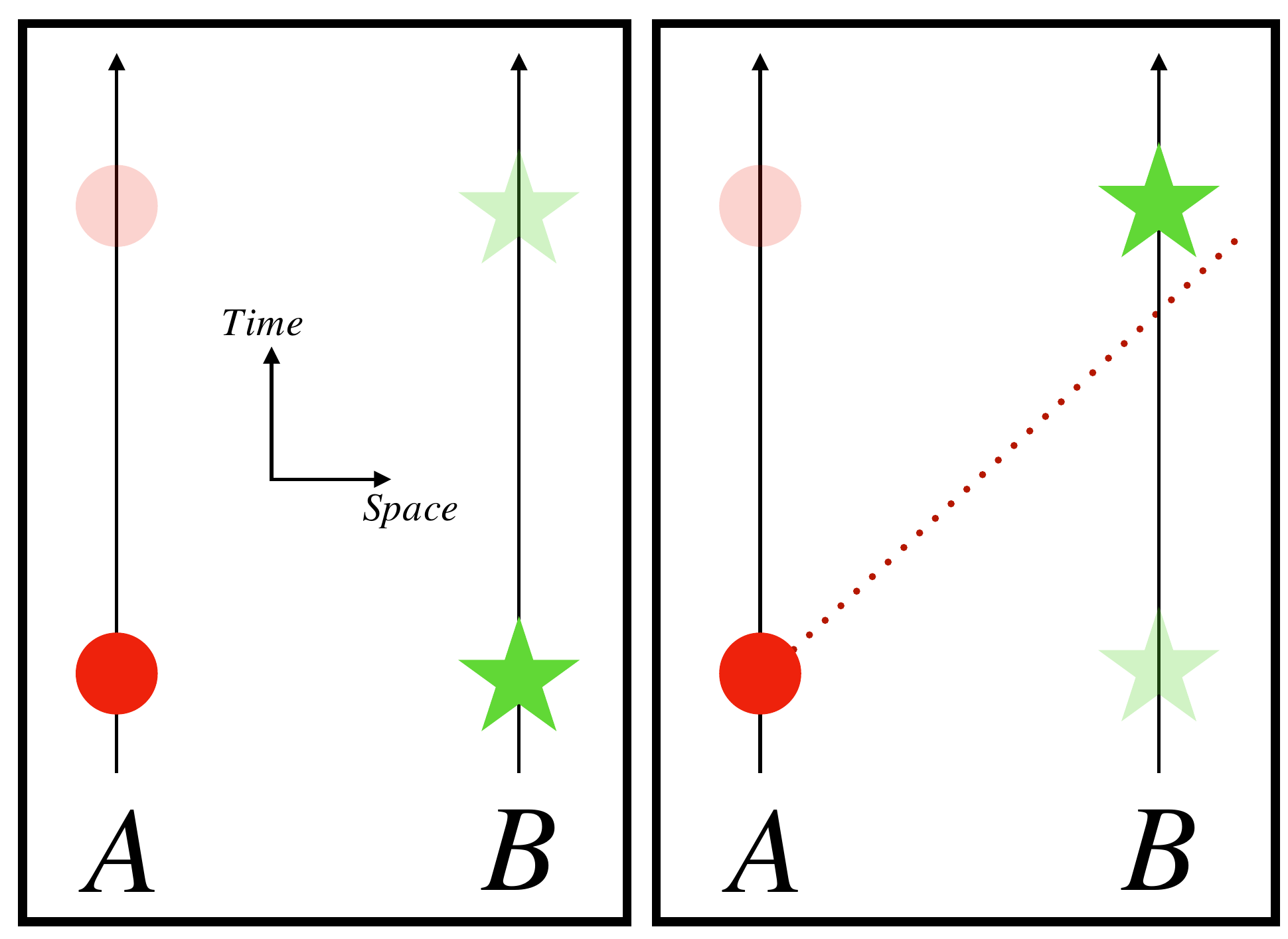}
  \caption{Left: scenario (i) in which the two detectors are in a superposition of being active either both in the past or both in the future, controlled by the state of the control qubit $\ket{0}$ or $\ket{1}$ respectively. In this case the
  spacetime
   activation  regions for the two detectors are spacelike in both branches of the quantum superposition. Right: scenario (ii) with indefinite causal order. In this case the two detectors are active when timelike related. The quantum superposition between ``$A$ before $B$'' and ``$B$ before $A$'' is controlled by the state of the control qubit $\ket{0}$ or $\ket{1}$ respectively. Note that in the main text, scenario (ii) also encompasses the case of space-like separated detectors where $A$ and $B$ are placed at different slices of constant time. (Colours online).}
  \label{fig1}
\end{figure}

We shall focus on two scenarios, depicted in Fig.~\ref{fig1}, which allow us to probe two distinct physical situations. Without loss of generality in all cases considered we assume a fixed but arbitrary  foliation of flat spacetime into space-like slices (hypersurfaces), each associated with a different time coordinate. Furthermore, in all scenarios each detector is switched on exactly once.
In the first scenario (i) detectors $A$ and $B$ are switched on jointly on a common space-like slice, but in a quantum-controlled superposition of two slices associated with different times. We call this scenario a \textit{past-future} (PF) superposition. The terms `past' and `future' refer to the relation between the superposed switching times.  
This case allows us to probe how entanglement between space-like regions of the field is affected by quantum delocalisation of these regions in time. 
In the second scenario (ii) detector $A$ is switched on on the past slice and $B$ on the future slice, in superposition with $A$ and $B$ interchanged.  We call this scenario a \textit{cause-effect} (CE) superposition, since for time-like separation between the involved regions the relation between $A$ and $B$ is causally indefinite -- we have a superposition of $A$ in the past causal cone of  $B$ and vice versa. Importantly, even for space-like separation between $A$ and $B$ the CE scenario is not equivalent to the PF scenario, which justifies making a general distinction between (i) and (ii) -- see Supplementary Materials~\cite{SI} for further discussion.

We begin by determining  the density matrix of the tripartite system $ABC$ by tracing out the field degree of freedom. The evolution of the entire system is described by a unitary map that can be expanded in a Dyson series to  second order in the coupling constant as $U=\mathbb{I}+U^{(1)}+U^{(2)}+\mathcal{O}(\lambda^3)$. More explicitly,
\begin{align}
    & U=\mathbb{I}-i\int_{-\infty}^{\infty}dt H_I(t) \nonumber\\
&\qquad    -\int_{-\infty}^{\infty}dt \int_{-\infty}^{t}dt'H_I(t)H_I(t')+\mathcal{O}(\lambda^3),
\end{align}
where $\mathcal{O}(\lambda^2)$ terms include terms $\lambda_A^2,\,\lambda_B^2,\,\rm{and}\,\lambda_A\lambda_B$. While here we employ a perturbative approach, it should be noted that some of the results in the following  can be reached also with non-perturbative methods (cf.~\cite{SI}).  

The initial state of $ABC$ and the field is chosen to be
\be
 \rho_0=\ket{0}_A \prescript{}{A}{\bra{0}}\otimes\ket{0}_B \prescript{}{B}{\bra{0}}\otimes \ket{+}_C 
 \prescript{}{C}{\bra{+}} \otimes\ket{0}_F \prescript{}{F}{\bra{0}},
\ee 
i.e., both detectors are in their ground states, the control qubit is in a coherent superposition of computational basis states, where $\ket{\pm}=(\ket{0}\pm\ket{1})/\sqrt{2}$, and the field is in the vacuum state. This choice of initial state greatly simplifies the computations, without concealing the physics of the problem, and it is instrumental to the study of the entanglement harvesting. 

The state after a time $\tau$ such that $\max[\rm{Supp}(\chi^D_i(t))]<\tau\,\forall i=\{0,1\},\,D=\{A,B\}$, is given by 
\begin{align}
\rho_f &= U\rho_0U^\dag =\rho_0+U^{(1)}\rho_0+\rho_0 U^{(1)\dag}+U^{(1)}\rho_0 U^{(1)\dag} \nonumber\\
& \qquad \qquad\qquad+\rho_0 U^{(2)\dag}+U^{(2)}\rho_0+\mathcal{O}(\lambda^3)
\end{align}
By tracing out the field, the reduced density matrix for the qubits $ABC$  --- to second  order in the coupling constants --- can be obtained as
\begin{align}
    & \rho_{ABC}=\rm{Tr}_F[\rho_f]  \nonumber\\
    &=\rm{Tr}_{F}[\rho_0+U^{(1)}\rho_0 U^{(1)\dag}+\rho_0 U^{(2)\dag}+U^{(2)}\rho_0]+\mathcal{O}(\lambda^4) 
    \nonumber \\
&=\frac{1}{2}\sum_{ij}
     \begin{bmatrix}
    1+Y_{ii}+Y_{jj}^* & 0 & 0 & \mathcal{M}_{jj}^* \\
     0 & P_{B,ij} & \mathcal{L}_{AB,ji}^* & 0 \\ 
    0 & \mathcal{L}_{AB,ij} &  P_{A,ij} & 0 \\
    \mathcal{M}_{ii} & 0 & 0 & 0 
\end{bmatrix}\otimes \ket{i}_C  \prescript{}{C}{\bra{j}}, 
\label{finalrho}
\end{align}
whose different elements are explicitly given in~\cite{SI}.

The terms $P_{D,ij}$ contain two-point function field correlations between the same spacetime point/region (corresponding to the position and time at which detector $D$ is active) if $i=j$; and two-point correlations between past and future activation regions of the same detector if $i\neq j$.  The  terms $P_{D,ii}$ are referred to as \textit{local} terms since they involve the ``self'' correlations between the local spacetime-region of activation of a single detector, whereas the terms $P_{D,i\neq j}$, though referring to a single detector,  are non-local since they contain correlations between different spacetime regions.The terms $Y_{ii}$ are similar in nature to the terms $P_{D,ii}$ and are related to them via $Y_{ii}+Y^*_{ii}=-(P_{A,ii}+P_{B,ii})$. 
The term $\mathcal{M}$ quantifies the non-local field correlations between different detectors and is responsible for the harvested entanglement from the field to the detectors. The terms $\mathcal{L}_{AB}$ are the leading order 
contributions to classical correlations between the detectors~\cite{Sachs:2017exo} and are the only terms that contribute to the mutual information between them.

The state of the control qubit ($C$) governs the choice of scenario.   Consider first scenario (i) illustrated in the left panel in Fig.~\ref{fig1}. We see that $\rm{Supp}(\chi_{A,i})=\rm{Supp(\chi_{B,i})}$ for $i=0,1$. Contrast this with  the right panel of Fig.~\ref{fig1} depicting scenario (ii), where now  $\rm{Supp}(\chi_{A,i})=\rm{Supp(\chi_{B,j})}$ for $i\neq j$. 
We eliminate the control qubit by either tracing it out or measuring it in  the basis $\{\ket{+},\ket{-}\}$. Our notation applies to both scenario by just a different labelling of the window functions.

Tracing out the control qubit yields
\begin{equation}
     \rho_{AB}^{(\rm{tr})}=\begin{bmatrix}
    1-(P_A^{(\rm{tr})}+P_B^{(\rm{tr})}) & 0 & 0 & \mathcal{M}* \\
     0 & P_B^{(\rm{tr})} & \mathcal{L}_{AB}^{(\rm{tr})*} & 0 \\ 
    0 & \mathcal{L}_{AB}^{(\rm{tr})}& P_A^{(\rm{tr})} & 0 \\
    \mathcal{M} & 0 & 0 & 0 
\end{bmatrix},
\label{rhoAB}
\end{equation}
where
\begin{align}
    & P_D^{(\rm{tr})}=\frac{1}{2}(P_{D,00}+P_{D,11}) \label{PD1}\\
    &\mathcal{L}_{AB}^{(\rm{tr})}=\frac{1}{2}(\mathcal{L}_{AB,00}+\mathcal{L}_{AB,11})\\
    &\mathcal{M}=\frac{1}{2}(\mathcal{M}_{00}+\mathcal{M}_{11}).
\end{align}
The results is a classical (probabilistic) mixture of \textit{two detectors active only once}, either spacelike related (i), or timelike related (ii), since the reduced state in Eq.~\eqref{rhoAB} is the mixture (with equal weights) of the classical case in which both detectors are active only once. This should have been expected. Indeed, borrowing some intuition from the process matrix formalism~\cite{2012NatCo...3.1092O}, the situation is similar to the \textit{quantum switch} in which an indefinite causal order is present only if the control qubit is allowed to quantum interfere~\cite{2013PhRvA..88b2318C,2015NatCo...6.7913P,Rubinoe1602589,2018PhRvL.121i0503G}. Tracing out the control qubit corresponds to losing the coherence of the superposition, effectively yielding `which-when' information about detector activation and thus remaining with a classical mixture. 

Conversely, measuring the control qubit in the $\{\ket{+},\ket{-}\}$ basis and retaining, e.g., the $\ket{+}$ results, allows the control qubit to interfere, yielding
\begin{equation}
    \rho_{AB}^{(+)}= \begin{bmatrix}
    1-(P_A^{(+)}+P_B^{(+)}) & 0 & 0 &\mathcal{M}* \\
     0 & P^{(+)}_{B} & \mathcal{L}_{AB}^{(+)*} & 0 \\ 
    0 & \mathcal{L}_{AB}^{(+)} &  P^{(+)}_{A} & 0 \\
    \mathcal{M} & 0 & 0 & 0 
\end{bmatrix}
\label{rhoABICO}
\end{equation}
where now
\begin{align}
    & P_D^{(+)}=\frac{1}{4}(P_{D,00}+P_{D,01}+P_{D,10}+P_{D,11}) \label{PD2}\\
    &\mathcal{L}_{AB}^{(+)}=\frac{1}{4}(\mathcal{L}_{AB,00}+\mathcal{L}_{AB,11}+\mathcal{L}_{AB,01}+\mathcal{L}_{AB,10}).
    \label{LAB13}
\end{align}
We see that, while the $\mathcal{M}$ terms are the same whether we trace out the control qubit or let it interfere, the other terms in the density matrix differ in the two cases. In particular, in $\rho_{AB}^{(+)}$ there appear terms $P_{D,i\neq j}$ and $\mathcal{L}_{AB,01(10)}$ that contain two-point correlation functions of the field between two different spacetime regions in which the same detector or both the detectors are  activated in superposition. These quantum interference terms are signatory of the coherence of the superposition. It should be noted that, without the controlled superposition the field correlations contained in the interference terms would not be accessible with a single use of each detector. By placing the detectors in quantum temporal superposition, we read out correlations between space time regions which can be accessed in the standard schemes with no superposition only if the detectors are used more than once (in more than one space-time region).

 \begin{figure*}[t!]
\centering
\includegraphics[width=0.32\textwidth]{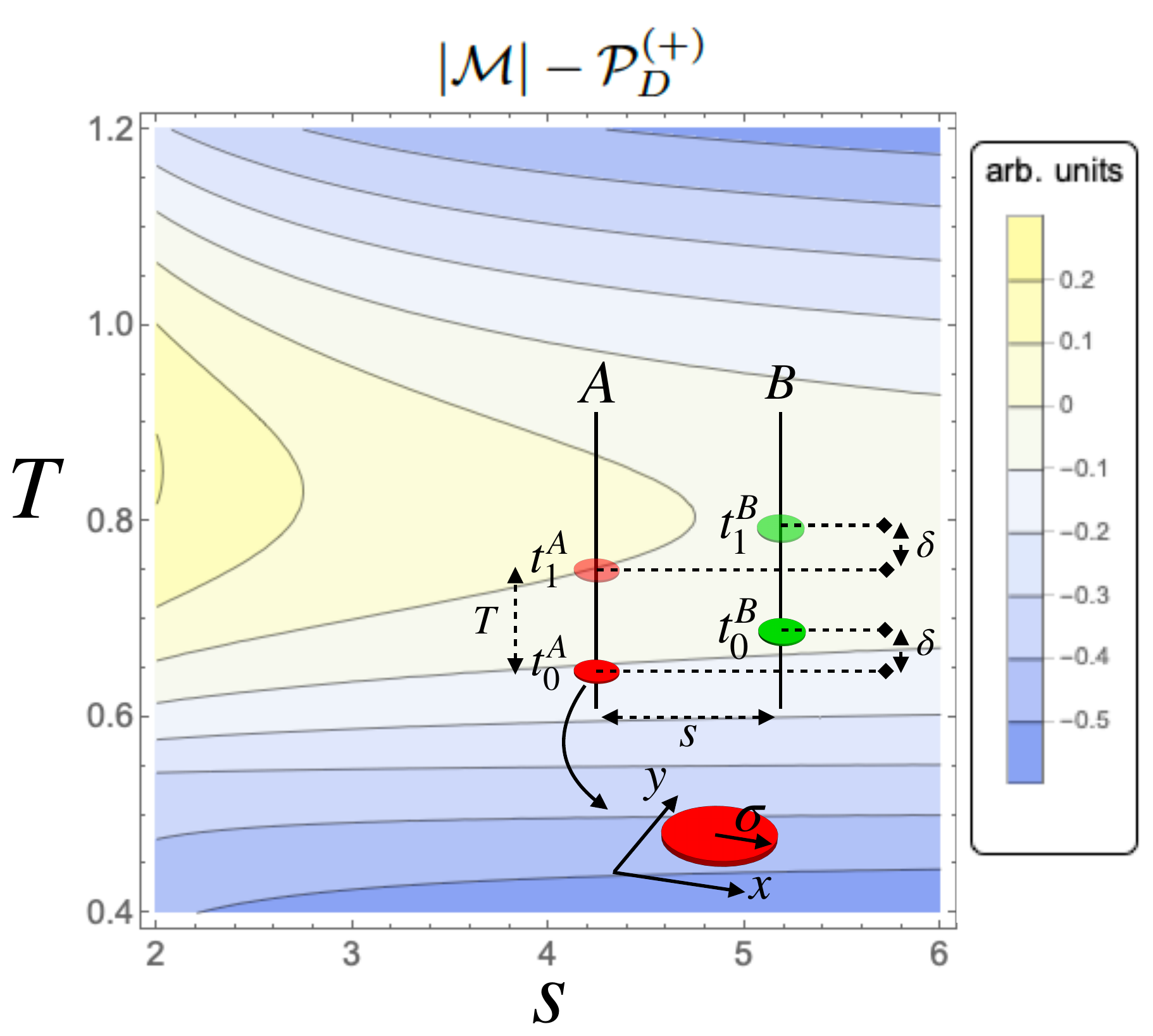}
\includegraphics[width=0.29\textwidth]{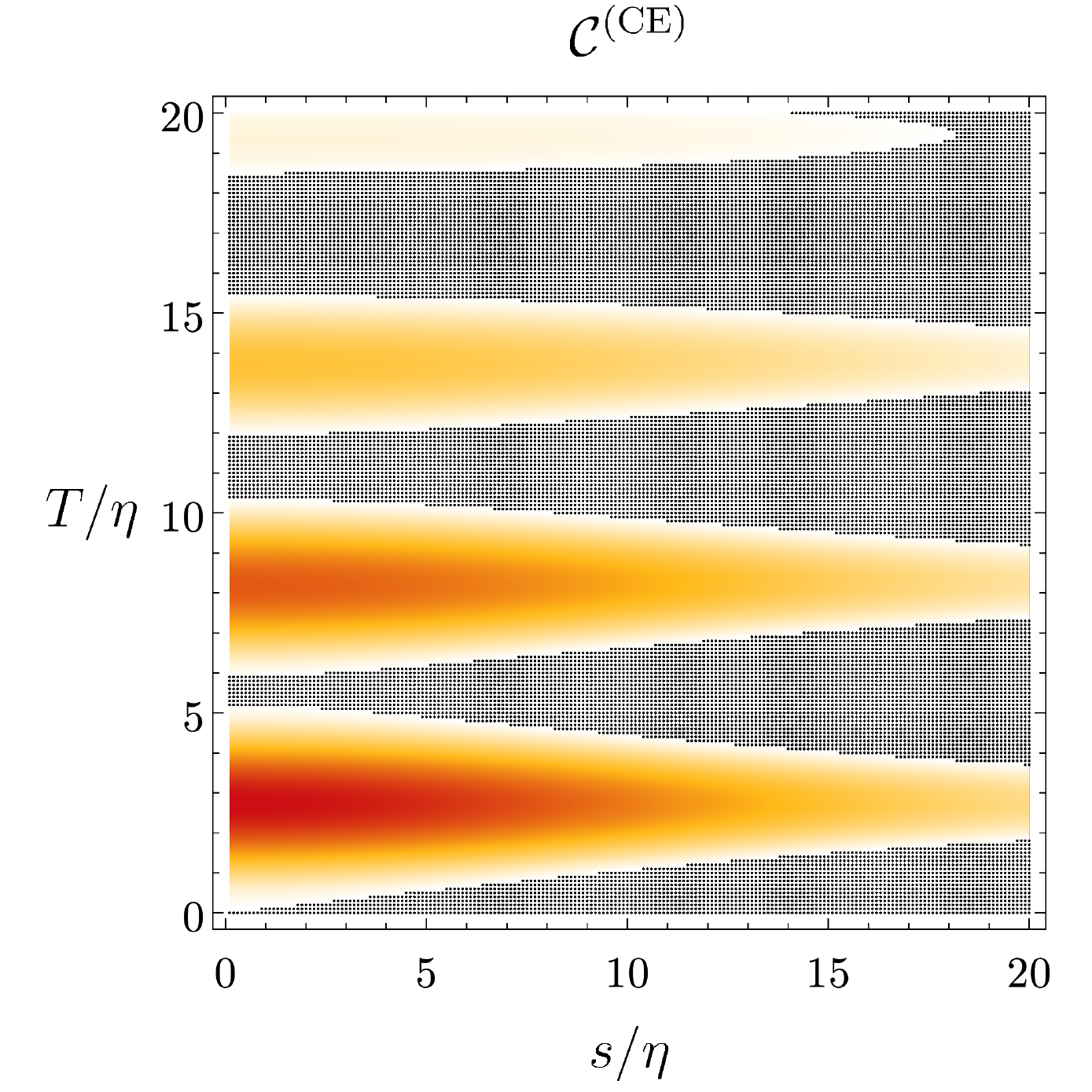}
\includegraphics[width=0.37\textwidth]{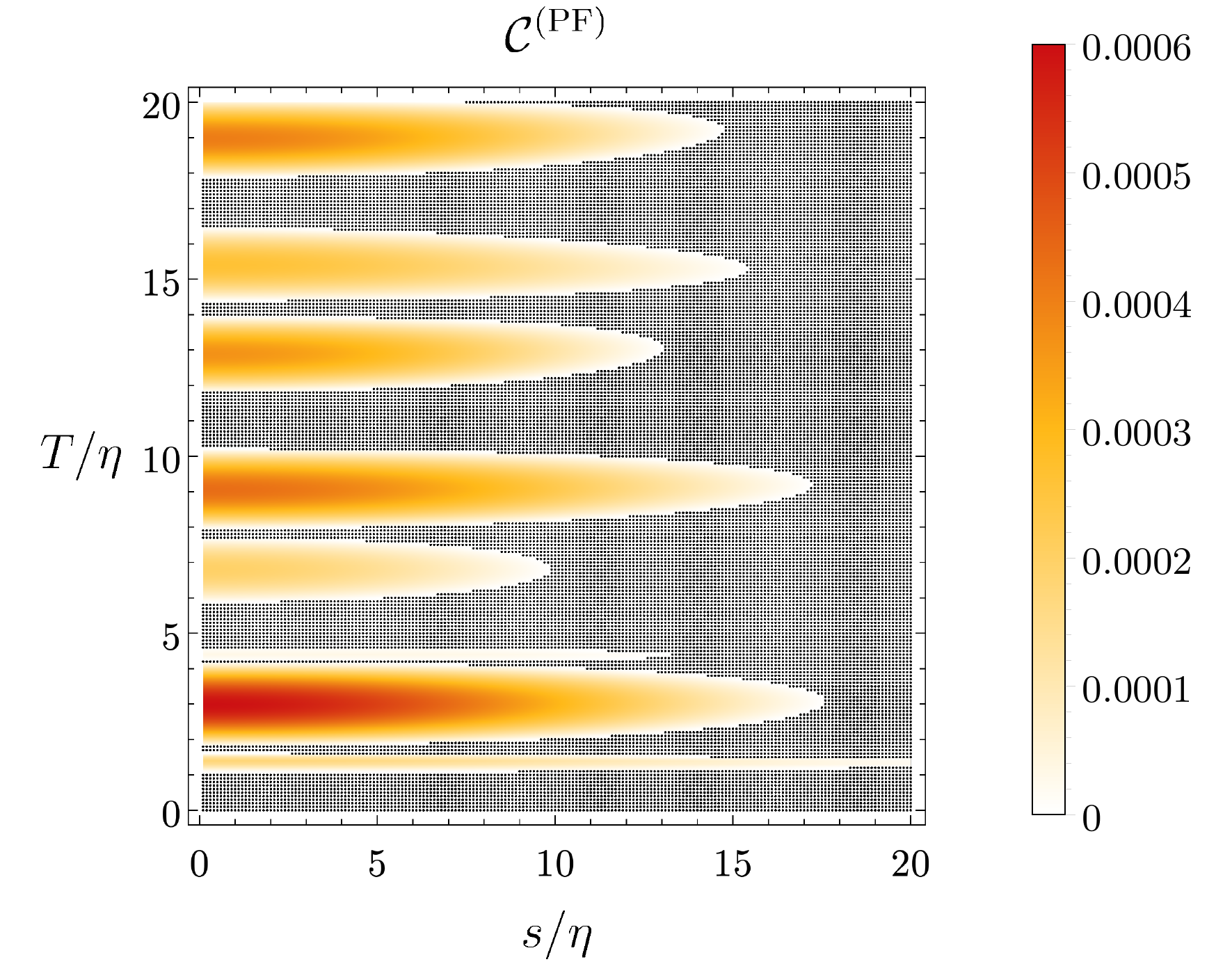}
  \caption{ Left: {perturbative} violation of the no-go theorem in 2+1 dimension for detectors with spatial compact support and Dirac-delta window function. The spacetime configuration of the two detectors --- corresponding to a PF scenario with detectors $A,B$ active at times $t^A_i$ and $t^B_i$ with $i=0,1$ determined by the control qubit --- is shown on the lower-right. The disks of radius $\sigma$ represent the spatial compact support. The violation happens whenever $|\mathcal{M}|-\mathcal{P}^{(+)}_{D}>0$. We plot this quantity as a function of spatial separation $s$ and temporal separation $T$ between activations in the different branches of the superposition. No violation occurs if $\mathcal{P}^{(+)}_{D}$ is replaced with $\mathcal{P}^{(\rm{tr})}_{D}$. Values used for the parameters are $\Omega=3,\,\sigma=1,\,\delta=10^{-5}$. Note that, when $T<s-2\sigma-\delta\, \land\, s>2\sigma\, \land\, \delta<s-2\sigma$ the activation regions of detector one are spacelike related with the ones of detector two. In the figure, this situation corresponds to the lower triangular area of the plot. 
  Centre: non-perturbative violation of the no-go theorem for Gaussian spatial profiles in 3+1 dimensions with CE superposition.   Right: non-perturbative violation of the no-go theorem for Gaussian spatial profiles in 3+1 dimensions with PF superposition.  In the centre and right panels black dots represent the regions where no entanglement can be harvested. The detectors have a window function $\chi(t) = \eta\delta(t-t_i^{D})$, a smearing function of $S(\bold{x}) = \frac{\exp\left[-(\bold{x}/9)^2\right]}{(9\sqrt{\pi})^3}$, an energy gap of $\Omega\eta=1$ and couple to the field with a strength of $\lambda=1$. The elements of the reduced density matrix have been numerically calculated to 15 significant figures. (Colors online). 
}
  \label{no-go}
\end{figure*}

This is our primary result: quantum controlled temporal superpositions allow UDW detectors to access field correlations between spacetime regions pertaining to the two different  branches in a quantum superposition. As we show in the following, this has relevant consequences for information-theoretic quantities,  such as the extractable entanglement from the quantum vacuum.
\\
\\
\noindent\emph{Entanglement  Harvesting --}
Let us now look at instances in which the additional correlations in~\eqref{rhoABICO}, stemming from the quantum interference, have relevant effects. 

We consider the process of entanglement harvesting from the quantum field vacuum, which is well-understood in flat spacetime~\cite{Salton:2014jaa,Martin-Martinez:2016}. Entanglement harvesting has attracted much interest for its foundational and applicative relevance ~\cite{Ralph:2015vra,Brown:2014pda,Salton:2014jaa,Martin-Martinez:2016,Rodriguez-Camargo:2016fbq,Richter:2017ljj,Richter:2017noq,Huang:2017yjt,Ardenghi:2018xrh}, with possible implications for the black hole information paradox and quantum gravity~\cite{Henderson:2017yuv,Pozas-Kerstjens:2017xjr,Ng:2018drz,Henderson:2018lcy,Ng:2018ilp,Sachs:2017exo,Simidzija:2018ddw,Trevison:2018cbf}.

Entanglement can be harvested from the quantum vacuum, in general, even for detectors that are spacelike separated~\cite{valentini1991non,Reznik2003}. However, the
procedure is subject to some strict limitations~\cite{Pozas-Kerstjens:2017xjr,PhysRevD.96.065008,PhysRevD.97.125002}. In particular, a no-go theorem has been proven~\cite{Pozas-Kerstjens:2017xjr,PhysRevD.96.065008} that states \textit{UDW detectors with Dirac-delta window functions, but arbitrary spatial profile and coupling strength, cannot harvest entanglement.} 

{Remarkably this no-go theorem is violated for quantum controlled temporal superposition of UDW detectors.} To prove this statement, we use as an entanglement measure the {concurrence} which, for identical detectors, is given by~\cite{PhysRevLett.80.2245}  \begin{equation}\label{conc}
\mathcal{C}^{(+/\rm{tr})}=  {2\max\left\{0,|\mathcal{M}|- \mathcal{P}^{(+/\rm{tr})}_{AB}
\right\} }
\end{equation}
in terms of the reduced density matrix elements in Eq.~\eqref{rhoABICO}, {where
$\mathcal{P}^{(+/\rm{tr})}_{AB} \equiv \sqrt{P^{(+/\rm{tr})}_{A} P^{(+/\rm{tr})}_{B} }$; if
the detectors are identical we write this as $\mathcal{P}^{(+/\rm{tr})}_{D}$.
}
 It is straightforward to show~\cite{PhysRevD.95.105009} that $\mathcal{C}^{(\rm{tr})} = 0$ since $|\mathcal{M}|\leq \mathcal{P}^{(\rm{tr})}_{AB}$ -- this is the no-go theorem.  However ${P}^{(+)}_D\leq  {P}^{(\rm{tr})}_D$ in general, and so $\mathcal{C}^{(+)}$ does not always vanish. {We illustrate this in Fig.~\ref{no-go}, where we see that  violations of the no-go theorem can be  proven 
 both perturbatively (left, using a compact spatial profile for each detector) and non-perturbatively (centre and right, using Gaussian spatial profiles).
These violations are due to the quantum interference} terms that contain the field correlations between the past and future activation regions of each detector. These correlations have the effect of lowering what can be interpreted as the detector excitation probability (the $P$ terms) with respect to a classical mixture~\eqref{rhoAB}. This in turn means that local noise effects are reduced by the quantum interference terms in such a way that entanglement harvesting is not precluded.

\begin{figure}[t!]
\centering
\includegraphics[width=0.43\columnwidth]{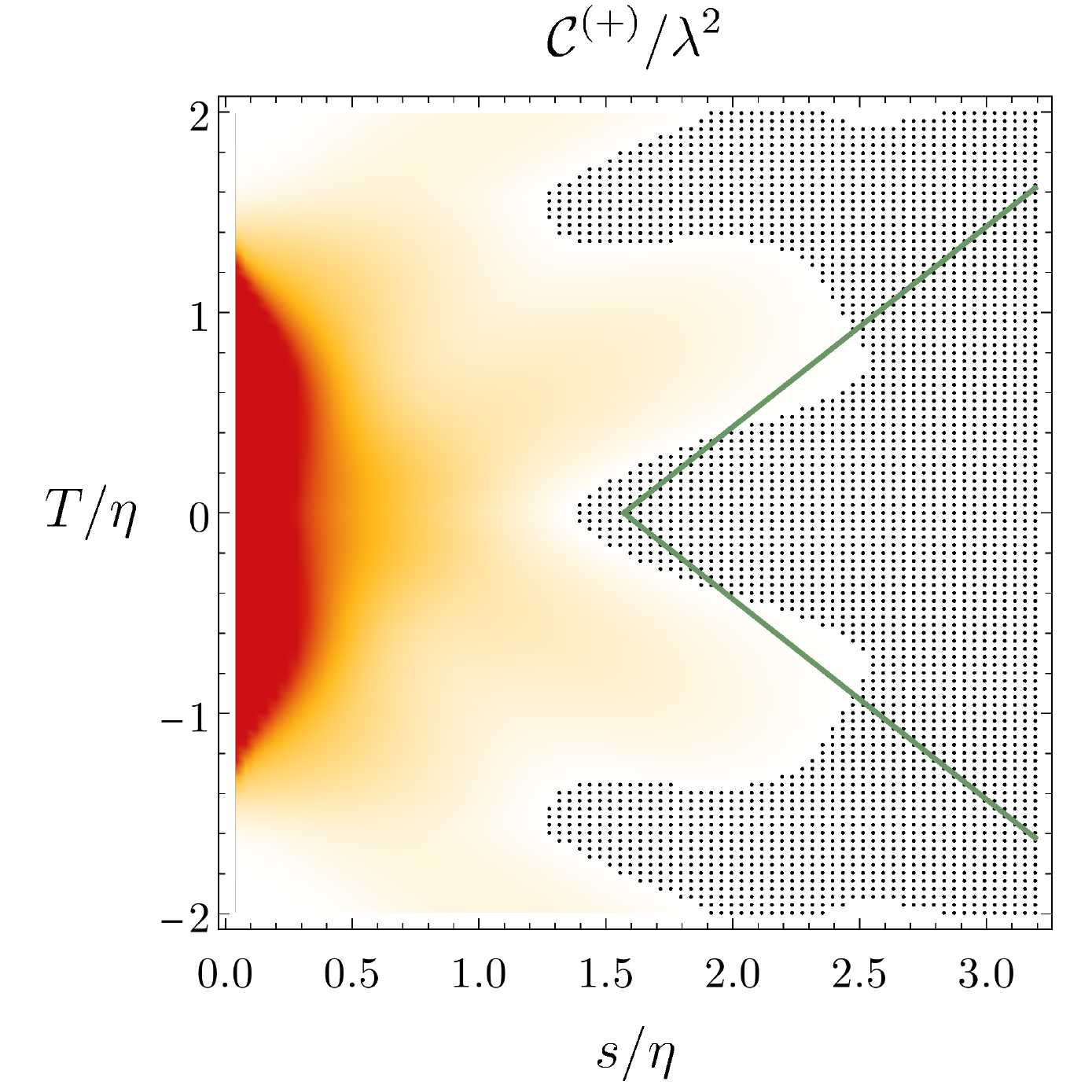}
\includegraphics[width=0.55\columnwidth]{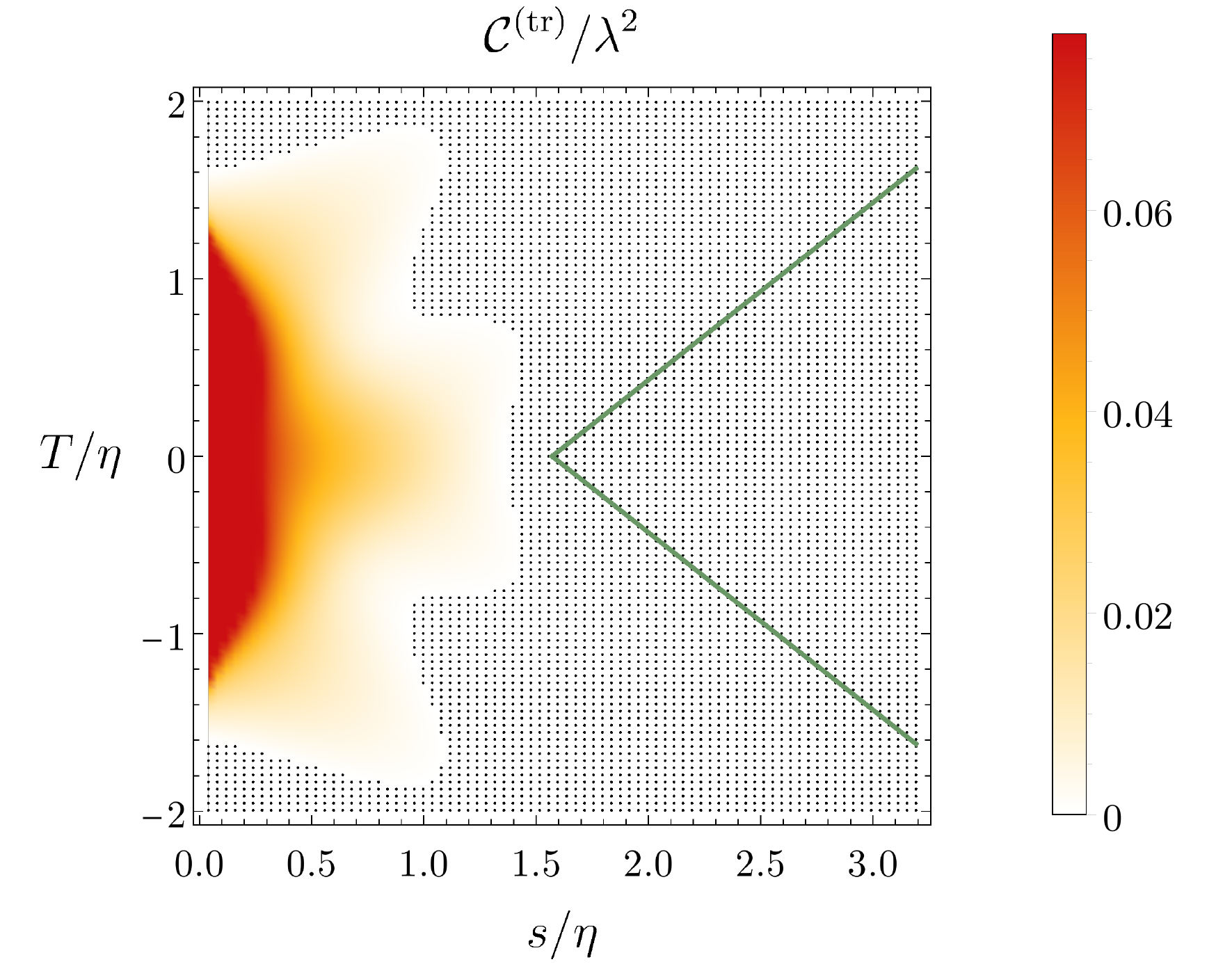}
  \caption{ %
  The concurrence of the perturbative reduced density matrix of two point-like detectors that couple to the field with a cosine window function in a (\textit{left}) quantum controlled CE superposition and (\textit{right}) a classical mixture of $A$ before $B$ and $B$ before $A$ plotted as a function of the spatial separation, $s$ and the temporal separation $T$ of the detectors.  The green lines mark the region of spacelike separation of the detectors and the black dots represent the regions where no entanglement can be harvested. For the chosen parameters, spacelike entanglement harvesting is possible only when the detectors are in the CE superposition.  The energy gap of the detectors is $\Omega\eta=3$.  The elements of the reduced density matrix have been numerically calculated to 15 significant figures. (Colors online).
  }
  \label{figCos}
\end{figure}

Another instance in which effects of   quantum controlled superposition are manifest is in  entanglement harvesting with pointlike detectors. In this case, in order to avoid divergences, the window function needs to be something other than a delta-function~\cite{Pozas-Kerstjens:2017xjr}.  In Fig.~\ref{figCos}, we employ a compact support window function, chosen to be $\cos(2t/\eta)$ for $-\pi/4 \leq t/\eta  \leq \pi/4$ and $0$ otherwise, $\eta$ being the ``width'' of such function. We see that  the quantum controlled superposition allows for spacelike entanglement harvesting whereas a classical mixture does not. Moreover, it is clear that whenever the two detectors are time-like related, ICO enhances the entanglement harvesting. This is a striking demonstration of how  quantum temporal interference effects can affect the ability of UDW detectors to harvest entanglement.
\\
\\
\noindent\emph{Discussion --}
 Indefinite casual order, or more generally, a quantum controlled superposition of detector activation times, affect  the  entanglement harvesting capabilities of UDW detectors  locally coupled to quantum fields.  This happens under a variety of scenarios,  producing the striking result that  spacelike entanglement harvesting becomes possible in situations not possible for classical mixtures.   {Note that the additional entanglement does not come directly from the control qubit. If there were no field (i.e.\ $\phi\to\mathbb{I}$) in Eq.~\eqref{Ham}, then after the interaction, the detectors would remain separable.}
 
While we have focused on entanglement harvesting as a relevant example, it should be noted that the quantum interference terms affect also the total amount of correlations that the detectors can harvest. This is quantified by their mutual information (cf.~\cite{SI}) in which  the term $\mathcal{L}_{AB}^{(+)}$ appears, which does not play any role in the entanglement harvesting, and which encodes field correlations between activation regions of the two detectors pertaining to separate quantum amplitudes of the process.

It is crucial to note that, the same field correlations entering Eq.~\eqref{PD2} would also be present if both detectors were active at two distinct times  (and in the same spacetime regions as for the quantum superposition case),  a scenario we call `double switching'. In fact, for double switching violations of the no-go theorem are  trivially possible, since the theorem requires  each detector being active only once \cite{PhysRevD.97.125002}.  However, the reduced state of $AB$ would differ from $\rho_{AB}^{(+)}$. Perturbatively, the difference would be in the $\mathcal{M}$ terms, meaning that the entanglement harvesting would be significantly different (while the mutual information would be the same).  Quantum temporal superposition is therefore  in principle distinguishable from the case in which each detector is twice activated (at least in certain parameter ranges, cf.~\cite{SI}), allowing for a clear  way to unambiguously state that the detectors in the superposition are activated only once.

In light of this discussion, our results imply that  UDW detectors with compact spacetime support qualify as good candidates  for \textit{local laboratories} in the process matrices framework~\cite{2012NatCo...3.1092O}. In this respect, it would be interesting to analyse coherent superpositions of detectors from the point of view of the process matrices formalism as a first step toward generalizing the formalism from quantum mechanics to quantum field theory. 

Finally, while correlations harvesting serves as just an instance of a more general framework, it well exemplifies the potential of probing a quantum field --- and thus spacetime itself~\cite{kempf2018quantum}, to some extent --- with quantum coherent superpositions of locally coupled detectors. This opens up to the possibility to use coherent superpositions of detectors as efficient probes of quantum fields. 
\\
\\
\emph{Acknowledgements --}
AB  acknowledges  the  hospitality  of  the  Institute  for  Theoretical  Physics  and  the  ``Nonequilibrium  quantum  dynamics''  group  at  Universit\"at  Stuttgart,  where  part  of  this  work was  carried  out. AB also  acknowledge  financial  support from H2020 through the MSCA IF pERFEcTO (GrantAgreement nr. 795782). 
\v{C}B and ECR acknowledge support from the research platform Testing Quantum and Gravity Interface with Single Photons (TURIS), the Austrian Science Fund (FWF) through the projects BeyondC (F7113-N48), and the doctoral program Complex Quantum Systems (CoQuS) under Project No. W1210-N25. They acknowledge financial support from the EU Collaborative Project TEQ (Grant Agreement No. 766900). \v{C}B acknowledge financial support from the Austrian-Serbian bilateral scientific cooperation no. 451-03-02141/2017-09/02, Foundational Questions Institute (FQXi) and the ID\# 61466 grant from the John Templeton Foundation, as part of the “The Quantum Information Structure of Spacetime (QISS)” Project (qiss.fr).. CB acknowledge support the Austrian Science Fund (FWF) through the projects ZK3 (Zukunftskolleg) and BeyondC (F7113-N48). This work was funded by a grant from the Foundational Questions Institute (FQXi) Fund. ECR is supported in part by the Program of Concerted Research Actions (ARC) of the Université libre de Bruxelles. MZ acknowledges funding from ARC through grants DECRA DE180101443 and EQuS CE170100009.  LJH and RBM acknowledge support from the Natural Sciences and Engineering Research Council of Canada.

 \bibliography{references2} 
 
 \newpage
\pagebreak
\widetext
\begin{center}
\textbf{\large Supplemental Materials: Quantum Temporal Superposition: the case of QFT}
\end{center}
\setcounter{equation}{0}
\setcounter{figure}{0}
\setcounter{table}{0}
\setcounter{page}{1}
\makeatletter
\renewcommand{\theequation}{S\arabic{equation}}
\renewcommand{\thefigure}{S\arabic{figure}}
\renewcommand{\bibnumfmt}[1]{[S#1]}
\renewcommand{\citenumfont}[1]{S#1}

In this Supplementary Material we derive the explicit expressions for the density matrices Eqs.~\eqref{finalrho}. We further specify these expressions to the case of $\delta$-window function with arbitrary spatial profile, i.e., the case of interest for the violations of the no-go theorem as reported in the main text. Moreover, we report also the reduced density matrix for the case of detectors each active two times.

\section{Detailed derivation of the density matrices in Eqs. (\ref{finalrho})}

We start from the initial state of $ABC$ and field that we chose as
\begin{equation}
    \rho_0=\frac{1}{2}\sum_{i,j}\ket{0}_A\bra{0}\otimes\ket{0}_B\bra{0}\otimes\ket{0}_F\bra{0}\otimes\ket{i}_C\bra{j},
    \label{eq:rho0}
\end{equation}
i.e., both detectors in the ground state, field in its vacuum state, and the control qubit in the $\ket{+}$ state. 

The state of $ABC$ at the final time is obtained by tracing out the field degree of freedom. For this we expand the time evolution operator to second order in the coupling constant(s) 
\begin{align}
    & U=\mathbb{I}-i\int_{-\infty}^{\infty}dt H_I(t)-\int_{-\infty}^{\infty}dt \int_{-\infty}^{t}dt'H_I(t)H_I(t')+\mathcal{O}(\lambda^3)
    \end{align}
The final state is given by $\rho_f=U\rho_0U^\dag$ and 
 taking  the trace over the field  yields \eqref{finalrho}
\begin{align}
       & \rho_{ABC} = Tr_F[\rho_f] =\frac{1}{2}\sum_{ij}
     \begin{bmatrix}
    1+Y_{ii}+Y_{jj}^* & 0 & 0 & \mathcal{M}_{jj}^* \\
     0 & P_{B,ij} & \mathcal{L}_{AB,ji}^* & 0 \\ 
    0 & \mathcal{L}_{AB,ij} &  P_{A,ij} & 0 \\
    \mathcal{M}_{ii} & 0 & 0 & 0 
\end{bmatrix}\otimes \ket{i}_C  \prescript{}{C}{\bra{j}} +\mathcal{O}(\lambda^4).
\end{align}
where the various elements are
 \begin{align}
       & P_{D,ij} = \lambda_D^2 \int_{-\infty}^{\infty}dt\int_{-\infty}^{\infty}dt'\chi_{D,i}(t)\chi_{D,j}(t')\mathbf{\langle\phi(x_{D}(t'))\phi(x_{D}(t))\rangle}   e^{i\Omega_A (t-t')} 
        \end{align}
where $D=A,B$ for each of detectors $A$ and $B$, and
 \begin{align}
       & \mathcal{L}_{AB,ij} = \lambda_A\lambda_B \int_{-\infty}^{\infty}dt\int_{-\infty}^{\infty}dt' \chi_{A,i}(t)\chi_{B,j}(t')\mathbf{\langle\phi(x_{B}(t'))\phi(x_{A}(t))\rangle} e^{i\Omega_A t}e^{-i\Omega_B t'} 
    \end{align}
    \begin{align}
    & Y_{ii} = - \sum_{D} \lambda_D^2 \int_{-\infty}^{\infty}dt\int_{-\infty}^{t}dt'\chi_{D,i}(t)\chi_{D,i}(t')\mathbf{\langle\phi(x_{D}(t))\phi(x_{D}(t'))\rangle}  e^{-i\Omega_D (t-t')}
\end{align}    
    \begin{align}
    & \mathcal{M}_{ii} = - \lambda_A\lambda_{B} \sum_{D\neq D'} \int_{-\infty}^{\infty}dt\int_{-\infty}^{t}dt'\chi_{D,i}(t)\chi_{D',i}(t')\mathbf{\langle\phi(x_{D}(t))\phi(x_{D'}(t'))\rangle} e^{i\Omega_D t}e^{i\Omega_{D'} t'}
\end{align}
and in all calculations we shall set $\lambda_A=\lambda_B$. 

\noindent Finally from the expressions above, it is easy to show that $Y_{ii}+Y^*_{ii}=-(P_{A,ii}+P_{B,ii})$.

\subsection{Entanglement measures:  {Entanglement of Formation} and Concurrence}

 One entanglement measure of a bipartite quantum state $\rho_{AB}$ is the entanglement of formation.  It has the operational interpretation as the number of Bell pairs required to prepare the state $\rho_{AB}$ using local operations and classical communication \cite{PhysRevA.54.3824}.  For two qubit detectors, it is given by the the equation \cite{PhysRevLett.80.2245}
\begin{equation}
  E_f(\rho_{AB}) = h\left(\frac{1+\sqrt{1-\mathcal{C}(\hat{\rho}_{AB})^2}}{2}\right) 
\end{equation}
where $h(x) = -x\log(x) -(1-x)\log(1-x)$\ and $\mathcal{C}(\hat{\rho}_{AB})$ is the concurrence, which itself is an entanglement monotone, defined as
\begin{equation}
  \mathcal{C}(\rho_{AB}) = \max\{0, w_1-w_2-w_3-w_4\}
\end{equation}
where the $w_i$'s are the square roots of the eigenvalues of $\rho_{AB}\big[(\sigma_y\otimes\sigma_y)\rho_{AB}(\sigma_y\otimes\sigma_y)\big]$ ordered from largest to smallest\cite{PhysRevLett.80.2245}.  For a density matrix of the form
\begin{equation}
  \rho_{AB} = \begin{pmatrix}
    \rho_{11} & 0 & 0 & \rho_{14} \\
    0 & \rho_{22} & \rho_{23} & 0 \\
    0 &\rho_{23}^* & \rho_{33} & 0 \\
    \rho_{14}^* & 0 & 0 & \rho_{44}
  \end{pmatrix}
  \label{eq:GeneralrhoAB}
\end{equation}
the concurrence is
\begin{equation}
  \mathcal{C}(\rho_{AB}) = 2\max\big\{0,|\rho_{14}|-\sqrt{\rho_{22}\rho_{33}},|\rho_{23}|-\sqrt{\rho_{11}\rho_{44}}|\big\}
\end{equation}
and for Eqs.~\eqref{rhoAB} and Eqs.~\eqref{rhoABICO}, the concurrence is
\begin{equation}
  \mathcal{C}\big(\rho_{AB}^{(+/\rm{tr})}\big) = 2\max\left\{0,|\mathcal{M}|- \sqrt{P^{(+/\rm{tr})}_AP^{(+/\rm{tr})}_B}\right\}.
\end{equation}

\subsection{Mutual Information}

The total amount of correlations between two systems is encoded by the mutual information, which for bipartite quantum system is given by the relative entropy between the state of the system and the tensor product of the reduced states of the subsystems, 
\begin{equation}
I(\rho_{AB})=S(\rho_{AB}||\rho_A\otimes\rho_B).
\end{equation} 

In the case of interest, the mutual information between the two UDW detectors is given perturbatively by
\begin{equation}
  I(\rho_{AB}^{(\cdot)})= \mathcal{L}_+ \log \mathcal{L}_+  + \mathcal{L}_- \log \mathcal{L}_-  - {P}_A^{(\cdot)} \log{P}_A^{(\cdot)} -{P}_B^{(\cdot)} \log{P}_B^{(\cdot)}
\end{equation}
where
\begin{equation}
 \mathcal{L}_\pm = \frac{1}{2}\left({P}_A^{(\cdot)}  + {P}_B^{(\cdot)}  \pm \sqrt{({P}_A^{(\cdot)} -{P}_B^{(\cdot)})^2+ 4 |\mathcal{L}_{AB}^{(\cdot)}|^2 } \right).
\end{equation}

It is interesting to note that, the expression for the mutual information of $AB$ for the case in which the control qubit is allowed to interfere, involves the term $\mathcal{L}_{AB}^{(+)}$ that encodes field correlations between activation regions of the two detectors not directly probed. Moreover, such terms do not play any role in the entanglement harvesting.

\section{$\delta$-window functions and violations of the no-go theorem}

 The $\delta$-window function given by 
\begin{equation}
  \chi_{D,i}(t) = \eta\delta(t-T_{D,i})
\end{equation}
allows us to calculate the final form of the detector-control system without the need of a perturbative expansion by following the method of~\cite{PhysRevD.96.065008}.  Time ordering requires that the switching order between detectors \textit{on both} branches of the superposition be fixed for the calculation, so we must consider the PF case separate from the CE case.

First, we consider the PF case under the assumption that $T_{A,0} \le T_{B,0} \le T_{A,1} \le T_{B,1}$. The time evolution operator can be simplified to
\begin{align}
  U 
  &= \exp\Big(-\iu H_{B,1}(T_{B,1}) \otimes \ket{1}_C\bra{1}\Big) \exp\Big(-\iu H_{A,1}(T_{A,1}) \otimes \ket{1}_C\bra{1}\Big) \exp\Big(-\iu H_{B,0}(T_{B,0}) \otimes \ket{0}_C\bra{0}\Big) \nonumber\\
  &\qquad \times \exp\Big(-\iu H_{A,0}(T_{A,0}) \otimes \ket{0}_C\bra{0}\Big) \nonumber\\
  &= U_0 \otimes \ket{0}_C\bra{0} + U_1 \otimes \ket{1}_C\bra{1}
\end{align}
where
\begin{align}
  U_i &= \Big(\mathbbm{1}_A \otimes \mathbbm{1}_B \otimes \cosh(X_{B,i})\cosh(X_{A,i})\Big) + \Big(\mathbbm{1}_A \otimes \mu_B(T_{B,i}) \otimes \sinh(X_{B,i})\cosh(X_{A,i})\Big) \nonumber\\
  &\qquad + \Big(\mu_A(T_{A,i}) \otimes \mathbbm{1}_B \otimes \cosh(X_{B,i})\sinh(X_{A,i})\Big) + \Big(\mu_A(T_{A,i}) \otimes \mu_B(T_{B,i}) \otimes \sinh(X_{B,i})\sinh(X_{A,i})\Big)
\end{align}
and
\begin{align}
  X_{D,i} = -\iu\lambda_D\eta \int d^3\bold{x} S_D(\bold{x}-\bold{x}_D)\phi(\bold{x},T_{D,i})
\end{align}
\begin{align}
  \mu_D(t) = (\ec^{\iu\Omega_D t}\sigma_+ + \ec^{-\iu\Omega_Dt}\sigma_-).
\end{align}
We consider the same initial state as Eqs.\eqref{eq:rho0}, and the final state is again $\rho_f=U\rho_0U^\dag$.  Tracing out the field degrees of freedom gives
\begin{align}
  \rho_{ABC}^{(\text{PF})} &= \frac{1}{32} \sum_{i,j}\Bigg\{\sum_{w,x,y,z} \Bigg[ \ec^{\frac{\iu}{2}\Omega_A[(1-z)T_{A,i}-(1-x)T_{A,j}] }\ec^{\frac{\iu}{2}\Omega_B[(1-y)T_{B,i}-(1-w)T_{B,j}]} \nonumber\\
  &\qquad\qquad \times f_{A,i}f_{A,j}f_{B,i}f_{B,j} \sum_{p,q,r,s} \Big(x^{(1+p)/2}w^{(1+q)/2}y^{(1-r)/2}x^{(1-s)/2} \nonumber\\
  &\qquad\qquad\qquad\qquad \times \ec^{pq(\frac{\iu}{2}\Theta_{A,j;B,j}+\omega_{A,j;B,j})} \ec^{pr(\frac{\iu}{2}\Theta_{A,j;B,i}+\omega_{A,j;B,i})} \ec^{ps(\frac{\iu}{2}\Theta_{A,j;A,i}+\omega_{A,j;A,i})} \nonumber\\
  &\qquad\qquad\qquad\qquad \times \ec^{qr(\frac{\iu}{2}\Theta_{B,j;B,i}+\omega_{B,j;B,i})} \ec^{-qs(\frac{\iu}{2}\Theta_{A,i;B,j}-\omega_{Ai;B,j})} \ec^{-rs(\frac{\iu}{2}\Theta_{A,i;B,i}-\omega_{A,i;B,i})} \Big) \nonumber\\
  &\qquad\qquad \times \ket{\frac{1-z}{2}}_A\bra{\frac{1-x}{2}} \otimes \ket{\frac{1-y}{2}}_B\bra{\frac{1-w}{2}}\Bigg] \otimes \ket{i}_C\bra{j}\Bigg\}
  \label{eq:deltaPF}
\end{align}
where $i,j\in\{0,1\}$ and $p,q,r,s,w,x,y,z \in \{+1,-1\}$.
\begin{align}
  f_{D,i} = \exp\left(-\frac{1}{2}\int d^3\bold{k}|\beta_{D,i}(\bold{k})|^2\right)
\end{align}
\begin{align}
  \Theta_{D,i;E,j} = \iu\int d^3\bold{k} \Big(\beta_{D,i}^*(\bold{k})\beta_{E,j}(\bold{k}) - \beta_{D,i}(\bold{k})\beta_{E,j}^*(\bold{k})\Big)
\end{align}
\begin{align}
  \omega_{D,i;E,j} = -\frac{1}{2}\int d^3\bold{k} \Big(\beta_{D,i}^*(\bold{k})\beta_{E,j}(\bold{k}) + \beta_{D,i}(\bold{k})\beta_{E,j}^*(\bold{k})\Big)
\end{align}
and
\begin{align}
  \beta_{D,i}(\bold{k}) = \frac{\iu\lambda_D\eta}{\sqrt{2|\bold{k}|}} \left(\int d^3\bold{x} S_D(\bold{x}) \ec^{\iu\bold{x}\cdot\bold{k}}\right) \ec^{-\iu(|\bold{k}|T_{D,i}-\bold{k}\cdot\bold{x}_D)}
\end{align}
for $D,E \in \{A,B\}$.

Next, we consider the CE scenario.  The time evolution operator is constructed in a similar way to the PF case, this time using the assumption that $T_{A,0} \le T_{B,1} \le T_{A,1} \le T_{B,0}$. The resulting detector-control density matrix for is
\begin{align}
  \rho_{ABC}^{(\text{CE})} &= \frac{1}{32} \sum_{i,j} \Bigg\{\sum_{w,x,y,z} \Bigg[\ec^{\frac{\iu}{2}\Omega_A[(1-iy-(1-i)z)T_{A,i}-(1-jw-(1-j)x)T_{A,j}]} \ec^{\frac{\iu}{2}\Omega_B[(1-(1-i)y-iz)T_{B,i}-(1-(1-j)w-jx)T_{B,j}} \nonumber\\
  & \qquad\qquad \times f_{D,j}f_{E,j}f_{F,i}f_{G,i} \sum_{p,q,r,s} \Big(x^{(1+p)/2}w^{(1+q)/2}y^{(1-r)/2}z^{(1-s)/2} \nonumber\\
  &\qquad\qquad\qquad\qquad \times \ec^{pq(\frac{\iu}{2}\Theta_{E,j;D,j}+\omega_{E,j;D,j})} \ec^{pr(\frac{\iu}{2}\Theta_{E,j;F,i}+\omega_{E,j;F,i})} \ec^{ps(\frac{\iu}{2}(\Theta_{E,j;G,i}+\omega_{E,j;G,i})} \nonumber\\
  &\qquad\qquad\qquad\qquad \times \ec^{qr(\frac{\iu}{2}\Theta_{D,j;F,i}+\omega_{D,j;F,i})} \ec^{qs(\frac{\iu}{2}\Theta_{D,j;G,i}+\omega_{D,j;G,i})} \ec^{rs(\frac{\iu}{2}\Theta_{F,i;G,i}+\omega_{F,i;G,i})} \Big) \nonumber\\
  &\qquad\qquad \ket{\frac{1-iy-(1-i)z}{2}}_A\bra{\frac{1-jw-(1-j)x}{2}} \otimes \ket{\frac{1-(1-i)y-iz}{2}}_B\bra{\frac{1-(1-j)w-jx}{2}}\Bigg] \nonumber \\
  &\qquad\qquad \otimes \ket{i}_C\bra{j}\Bigg\}.
  \label{eq:deltaCE}
\end{align}
The labels $D,E,F,G$ depend on the values of $i$ and $j$ and are summarised in Eq.~\eqref{eq:deltaCEDEFG} below.
\begin{align}
  &D = \begin{cases}
    B, &j=0\\
    A, &j=1
  \end{cases}
  &E = \begin{cases}
    A, &j=0\\
    B, &j=1
  \end{cases} \nonumber\\
    &F = \begin{cases}
    B, &i=0\\
    A, &i=1
  \end{cases}
  &G = \begin{cases}
    A, &i=0\\
    B, &i=1
  \end{cases}
  \label{eq:deltaCEDEFG}
\end{align}

 Once the control qubit is measured (or traced out), the density matrices given in Eqs.~\eqref{eq:deltaCE} and Eqs.~\eqref{eq:deltaPF} will take the form of Eq.~\eqref{eq:GeneralrhoAB} and are then used to compute the entanglement between detectors $A$ and $B$ shown in the centre and right figures of Fig.~\ref{no-go}.

\subsection{Perturbative expressions}
While the non-perturbative method discussed above is valid in general, the final expressions are quite cumbersome and the numerical analysis with spatial profiles other than Gaussian functions proves to be hard. 

In order to obtain expressions more amenable to a numerical analysis, and be able to investigate the case in which the two detectors have compact spatial supports, we resort here to a perturbative calculation. It should be noted that, the case of compact spatial support is particularly relevant since  only in this case we can, strictly speaking, consider spacelike separated detectors. 

\begin{figure}[h!]
\centering
\includegraphics[width=0.5\textwidth]{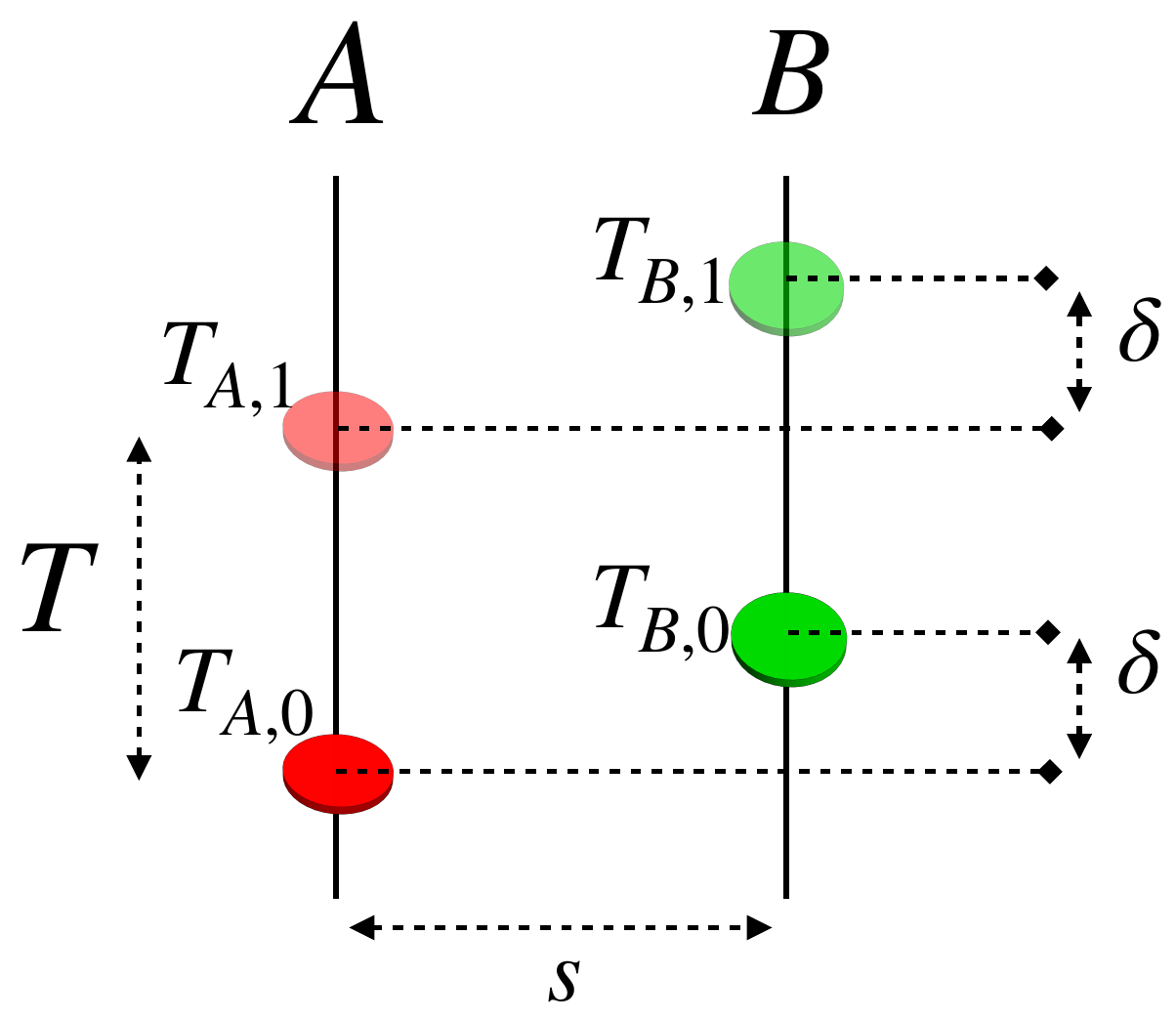}
  \caption{Spacetime configuration of the two UDW detectors in quantum superposition considered in Eqs.\eqref{Supp_exp}. }  
  \label{SuppFig_setup}
\end{figure}

The expressions in this section, when not otherwise stated, are valid for $\delta$-window functions detectors with arbitrary spatial profile in $n+1$ dimensions. We consider always identical detectors with window function $ \chi_{D,i}(t) = \delta(t-T_{D,i})$ and spatially separated by a distance $s$ in their mutual rest frame. In particular, we consider the case of scenario (i) with $T_{B,0(1)}=T_{A,0(1)}+\delta$ and $T_{D,1}-T_{D,0}=T>0$ for $D=A,B$, (cf. Fig.\ref{SuppFig_setup} and the spacetime configuration in the lower-right corner of the left figure in Fig.\ref{no-go}).

In this case, the relevant terms to consider for the violation of the no-go theorem are
\begin{align}\label{Supp_exp}
 & P_{D,ii}=\lambda^2\int d^n\mathbf{k}\frac{|\tilde{S}(\mathbf{k})|^2}{2|\mathbf{k}|}\\
 & P_{D,01}+P_{D,10}=2\lambda^2\int d^n\mathbf{k}\frac{|\tilde{S}(\mathbf{k})|^2}{2|\mathbf{k}|}\cos((|\mathbf{k}|+\Omega)T)\\
 & |\mathcal{M}|=\lambda^2\left|\int d^n\mathbf{k}\frac{|\tilde{S}(\mathbf{k})|^2}{2|\mathbf{k}|}e^{-i\mathbf{k}\cdot (\mathbf{x}_A-\mathbf{x}_B)}e^{-i|\mathbf{k}|\delta}e^{i\Omega (T_{A,0}+T_{B,0}+T)}\cos(\Omega T)\right|.
\end{align}
These are the terms that enter in the expression for the concurrence of the state of $AB$ (Eq.\eqref{conc}) and thus they characterize the entanglement harvesting at the perturbative level. 

\section{The double-switch scenario}
In this section we consider the case in which each detector is active two-times and no quantum control is present. The interaction Hamiltonian is given by 
\begin{equation}
    H_I(t)= \frac{1}{2}\sum_{D=A,B}\lambda_D(\chi_{D,0}(t)+\chi_{D,1}(t))(e^{i\Omega_D t}\sigma_+ +e^{-i\Omega_D t}\sigma_{-})\otimes\phi(x_{D}(t)),
\end{equation}
where the window functions is a combination of the window functions used in the main text in the quantum superposition case. Once the field is traced out, the density matrix of $AB$ at the final time is given by
\begin{equation}
   \rho^{(DS)}_{AB}= \begin{bmatrix}
    1-(P_A^{(+)}+P_B^{(+)}) & 0 & 0 & \mathcal{M}^{(DS)*} \\
     0 & P^{(+)}_{B} & \mathcal{L}_{AB}^{(+)*} & 0 \\ 
    0 & \mathcal{L}_{AB}^{(+)} &  P^{(+)}_{A} & 0 \\
    \mathcal{M}^{(DS)} & 0 & 0 & 0 
\end{bmatrix},
\end{equation}
where
\begin{align}
    & P_D^{(+)} =\frac{1}{4}(P_{D,00}+P_{D,01}+P_{D,10}+P_{D,11} )\\
    &\mathcal{L}_{AB}^{(+)}=\frac{1}{4}(\mathcal{L}_{AB,00}+\mathcal{L}_{AB,11}+\mathcal{L}_{AB,01}+\mathcal{L}_{AB,10})\\
    &\mathcal{M}^{(DS)}=\frac{1}{4}(\mathcal{M}_{00}+\mathcal{M}_{11}+\mathcal{M}_{01}+\mathcal{M}_{10}) \label{eq:MDS}
\end{align}
We see that (i) the density matrix contains the field correlations between all the four spacetime regions in which the detectors are active, analogously to the case in which the control qubit is measured in $\ket{+}$. However, in this case this is due to the fact that the detectors are actually active twice each, in contrast to the quantum case in  which each detector (in a quantum superposition) is used only once.  (ii) The $(1,4), (4,1)$ terms of the density matrix are different from the corresponding ones in Eq.\eqref{rhoABICO}.   Thus, the double-switch  scenario is  distinguishable  from the quantum coherent one already at the perturbative level.

 If a particular case of superposition is chosen (say PF), then it can be shown that if the $(1,4),(4,1)$ terms in the quantum case are equal to $2(\mathcal{M}_{00}+\mathcal{M}_{11})$ in the double-switch scenario, then the remaining cross terms, $2(\mathcal{M}_{01}+\mathcal{M}_{10})$, will be equal to the $(1,4),(4,1)$ terms in the other case of superposition (in this case CE) with the same time delay.  Mathematically
we have
\begin{equation}
  \rho_{AB}^{(\text{DS})} = \frac{1}{2}\left(\rho_{AB}^{(+_{\text{PF}})}+\rho_{AB}^{(+_{\text{CE}})}\right) 
\end{equation}
indicating that the double-switch scenario can be decomposed into the two quantum cases we considered.

\section{Inequivalence between scenario I and II}

 One might think that scenario (ii) when $A$ and $B$ are space-like separated is equivalent to scenario (i), where $A$ and $B$ are switched on simultaneously -- since for space-like separated regions one can find coordinates in which the switching is simultaneous. Despite this fact, there is no equivalence between these scenarios: in (i) the switching is simultaneous in each superposed amplitude (albeit at different times) whereas choosing coordinates in scenario (ii) so that $A$ and $B$ are simultaneous for one amplitude will result in $A$ being  before or after $B$ in the other amplitude, illustrated in the Supplementary Figure \ref{SuppFig_PFCE}, left panel. In other words, $A, B$ are simultaneous in superposition with $A$ earlier than $B$ (or $B$ earlier than $A$). An analogous difference arises if we try to change coordinates in scenario (i) so as to map it to scenario (ii). Using coordinates in which $A$ from one amplitude is simultaneous with $B$ from the other amplitude -- as it is in case (ii) -- actually yields a different scenario: where in each amplitude $A$ is switched on before $B$ (or $B$ before $A$ -- depending which amplitude is chosen to define simultaneity) -- see also Supplementary Figure \ref{SuppFig_PFCE}, right panel. In other words, we end up with $A$ before $B$ in superposition with $A$ before $B$ but both switched on some time later (or $B$ before $A$ in each superposed amplitude).  Recall, however that in scenario (ii) we have a superposition of $A$ switched on before $B$ and $B$ before $A$. Thus, even in case of space-like separation between the relevant regions, the two scenarios (i) and (ii) are not equivalent -- there is no single coordinate transformation that allows one to map them onto each other.
\begin{figure}[h!]
\centering
\includegraphics[width=0.9\textwidth]{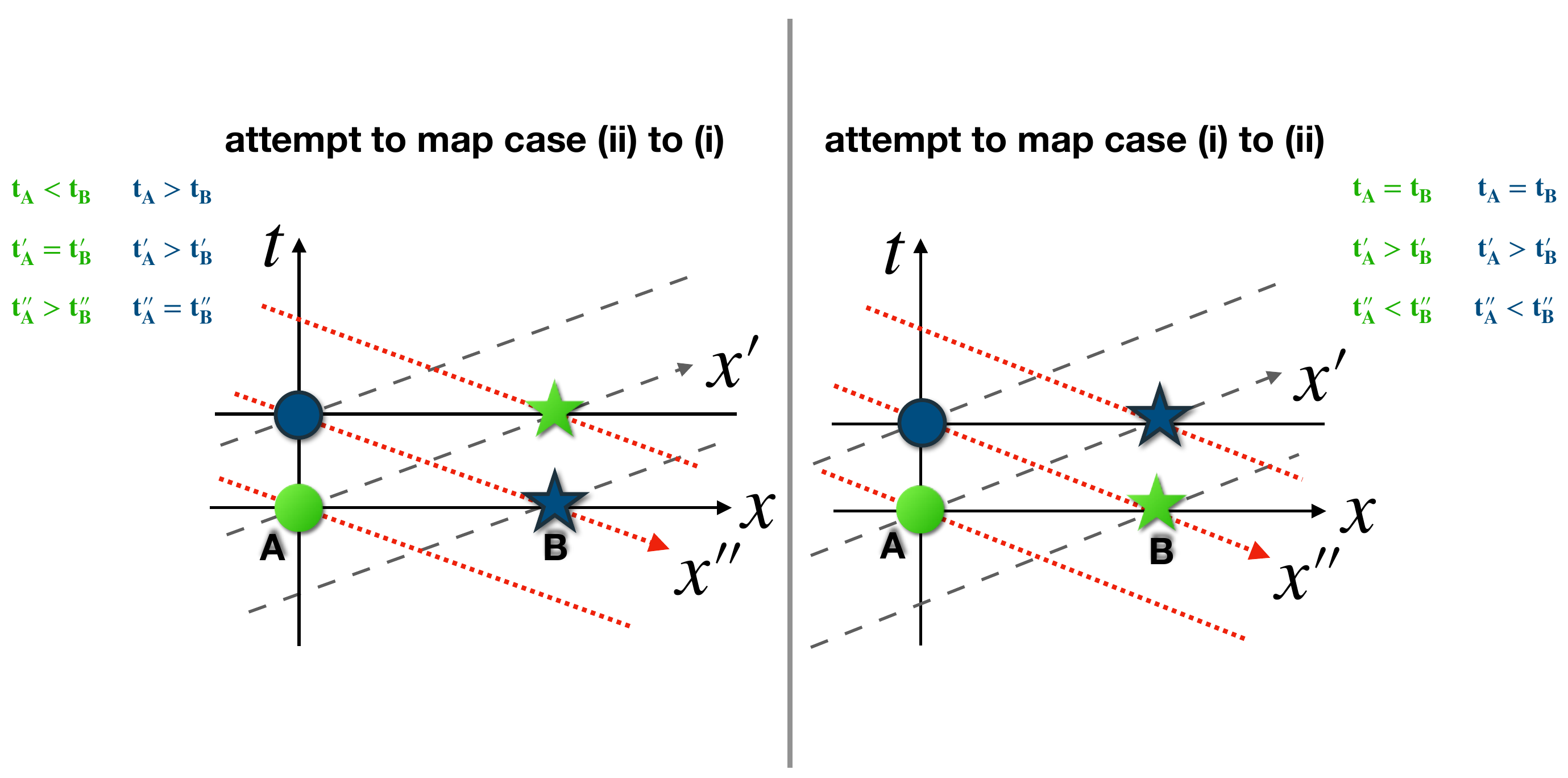}
  \caption{ It is not possible to map scenarios (ii) and (i) to each other. Left panel: when trying to map scenario (ii) to (i) we can  make make $A$ and $B$ simultaneous only in one of the superposed amplitudes -- rather than in each of them, as required in scenario (i). Right panel: when trying to map scenario (i) to (ii) we can  make make $A$ from one amplitude simultaneous $B$ from the other amplitude, but we do not end up with a superposition of $A$ earlier than $B$ and $B$ than $A$ -- in each amplitude $A$ will be  switched on earlier than $B$ (or in each amplitude $B$ will be earlier than $A$). In both panels, $x$ coordinates (black solid lines) make past $A$ and past $B$ simultaneous and future $A$ and future $B$ simultaneous, the $x'$ coordinates (grey dashed lines) make past $A$ simultaneous with future $B$ and the $x''$ coordinates (red dotted lines) make future $A$ simultaneous with past $B$.}
    \label{SuppFig_PFCE}
\end{figure}
 
 \end{document}